\documentclass[preprint]{aastex}
\usepackage{graphicx}
\usepackage{amssymb}
\usepackage{url}

\shorttitle{The Extended Emission of AXP 1E~1547.0--5408}
\shortauthors{Olausen et al.}

\begin{document}

\title{On the Extended Emission Around the Anomalous X-ray Pulsar 1E~1547.0--5408}

\author{S. A. Olausen\altaffilmark{1}, V. M. Kaspi\altaffilmark{1}, C.-Y. Ng\altaffilmark{1}, W. W. Zhu\altaffilmark{1}, R. Dib\altaffilmark{1}, F. P. Gavriil\altaffilmark{2,3}, and P. M. Woods\altaffilmark{4,5}}

\altaffiltext{1}{Department of Physics, Rutherford Physics Building, McGill University, 3600 University Street, Montreal, Quebec H3A 2T8, Canada}

\altaffiltext{2}{NASA Goddard Space Flight Center, Astrophysics Science Division, Code 662, Greenbelt, MD 20771, USA}

\altaffiltext{3}{Center for Research and Exploration in Space Science and Technology, University of Maryland Baltimore County, 1000 Hilltop Circle, Baltimore, MD 21250, USA}

\altaffiltext{4}{Dynetics, Inc., 1000 Explorer Boulevard, Huntsville, AL 35806, USA}

\altaffiltext{5}{Corvid Technologies, 689 Discovery Drive, Huntsville, AL 35806, USA}

\begin{abstract}
We present an analysis of the extended emission around the anomalous
X-ray pulsar 1E 1547.0$-$5408 using four \textit{XMM-Newton} observations
taken with the source in varying states of outburst as well as in
quiescence. We find that the extended emission flux is highly variable
and strongly correlated with the flux of the magnetar. Based on this
result, as well as on spectral and energetic considerations, we conclude
that the extended emission is dominated by a dust-scattering halo
and not a pulsar wind nebula (PWN), as has been previously argued.
We obtain an upper limit on the 2--10\,keV flux of a possible PWN
of $4.7\times10^{-14}\,\mathrm{erg\, s^{-1}\, cm^{-2}}$, three times
less than the previously claimed value, implying an efficiency for
conversion of spin-down energy into nebular luminosity of $<$$9\times10^{-4}$
(assuming a distance of 4\,kpc). We do, however, find strong evidence
for X-ray emission from the supernova remnant shell surrounding the
pulsar, as previously reported.
\end{abstract}


\section{Introduction}

Anomalous X-ray pulsars (AXPs) and soft gamma repeaters (SGRs) are
generally accepted as belonging to the class of neutron stars known
as `magnetars.' These sources are characterized by long (2--12\,s)
spin periods and spin-down rates that imply extremely high surface
dipole magnetic fields of $10^{14}\textrm{--}10^{15}\,\mathrm{G}$
(see \citealt{wt06} and \citealt{m08} for reviews). It is thought
that their large X-ray luminosities are powered by the decay of these
powerful magnetic fields, and fracturing of the crust and reconfiguration
of the magnetic field lines produce the glitches, bursts, and X-ray
variability seen in these sources \citep{td95,td96,tlk02}.

One important open question regarding AXPs and SGRs is whether they
produce particle outflows akin to those seen in conventional rotation-powered
pulsars. The latter are well known to produce often spectacular `pulsar
wind nebulae' (PWNe), the classic example of which is the Crab Nebula
\citep[see, e.g.,][for reviews]{krh06,gs06,kp08}. Such nebulae, evident
particularly at radio and X-ray energies, are the result of synchrotron
emission due to pulsar-produced relativistic electrons and positrons,
as they spiral in the ambient magnetic field. Given the large magnetic
energy reservoir hypothesized to exist in AXPs and SGRs, particle
outflows seem reasonable to consider, and indeed have been suggested
to be present ubiquitously in magnetars, either in continuous or sporadic
forms \citep{tb98,hck99}. This idea was initially buoyed by the claim
of an apparent wind nebula associated with SGR 1806$-$20 \citep{mtk+94}.
Although the latter association was later disproved \citep{hkc+99},
the possibility of a nebula-producing magnetar wind has not been.
Extended radio emission was unambiguously identified following one
SGR giant flare \citep{gkg+05,tgg+05,gle+05,fmg+06}, but is thought
to be from relativistic, weakly baryon-loaded magnetic clouds \citep{l06}
or from a baryonic outflow \citep{gle+05,g07}, and associated exclusively
with the flare. Recently \citet{rmg+09} have suggested that an unusual
X-ray nebula surrounding the relatively high-$B$ Rotating Radio Transient
(RRAT) J1819$-$1458 is magnetically powered, though the mechanism
for this is unclear.

The X-ray source 1E 1547.0$-$5408 was discovered with the \textit{Einstein}
X-ray satellite in 1980 by \citet{lm81}, but only recently was it
suggested to be a magnetar candidate on the basis of its spectrum,
variable nature, and likely association with the supernova remnant (SNR)
shell G327.24--0.13 \citep{gg07}. Radio observations of the source by
\citet{crhr07} revealed pulsations at a period of 2.1\,s and the measured
spin-down rate implied a surface magnetic field strength of $2.2\times10^{14}\,\mathrm{G}$
and a spin-down luminosity of $\dot{E}=1.0\times10^{35}\,\mathrm{erg\, s^{-1}}$.
1E 1547.0$-$5408 is thus the fastest-rotating and highest-$\dot{E}$
magnetar yet known.%
\footnote{See the McGill SGR/AXP Online Catalog at \url{http://www.physics.mcgill.ca/~pulsar/magnetar/main.html}.%
} Although observations in 2006 showed the source to be in quiescence,
a 2007 \textit{XMM-Newton} observation showed it to be in a high, apparently
post-outburst state \citep{hgr+08}. In 2008 October and again on
2009 January 29, 1E 1547.0$-$5408 underwent strong outburst events,
experiencing dramatic increases in its X-ray luminosity and exhibiting
many SGR-like bursts within a few hours. For further details on these
outbursts and the source's history, see \citet{ier+10}, \citet{kgk+10},
\citet{nkd+11}, \citet{bis+11}, \citet{sk11}, and \citet{dkg11}.

In observations of 1E 1547.0$-$5408 taken with \textit{Swift} and \textit{XMM-Newton}
following the 2009 outburst, \citet{tve+10} observed dust-scattering
X-ray rings centered on the magnetar and derived from them a source
distance of $\sim$4--5\,kpc. They also found evidence of time variable
diffuse emission around the source which they attributed to dust scattering
of the bursts and of the persistant X-ray emission.

Meanwhile, \citet[hereafter VB09]{vb09}, analyzing \textit{Chandra}
and \textit{XMM-Newton} observations of the source taken in 2006 when
it was in quiescence, detected extended emission and characterized
it as the result of a PWN, in analogy with those seen around rotation-powered
pulsars. They argued for a PWN based on the high flux level of the
extended emission, and because it appeared to have a harder spectrum
than the point source. This would make 1E 1547.0$-$5408 unusual among
the known magnetars, as no other such source has been shown to power
such emission. They also showed the presence of extended X-ray emission
coincident with the SNR shell.

Here we present an analysis of the extended emission around 1E 1547.0$-$5408
using multi-epoch \textit{XMM} data, in which the source flux varies
strongly, as does the putative nebular emission. We show conclusively
that the putative PWN reported by VB09 is in fact dominated by dust
scattering, rather than by emission from any pulsar outflow.

\section{Observations}

We obtained an observation of 1E 1547.0$-$5408 with the \textit{XMM-Newton
Observatory} on 2010 February 10 in order to track the star's properties
as it decayed to quiescence after the 2009 outburst. We also reanalyzed
three archival \textit{XMM-Newton} observations of the source: one in
2006 with the source in quiescence, the 2007 post-outburst observation,
and one taken 2 weeks after the 2009 outburst. For each observation,
the data from the two EPIC MOS cameras \citep{taa+01} were not suitable
for our analysis, either because the operating mode provided too small
a field of view or because the data were highly piled-up. We therefore
restricted our analysis to data from the EPIC pn camera \citep{sbd+01},
which had no such issues. The data from all four observations were
analyzed using the \textit{XMM-Newton} Science Analysis System (SAS)
version 10.0.2%
\footnote{See \url{http://xmm.esac.esa.int/sas/10.0.0/} %
} with calibrations updated 2010 July 29. Each observation was filtered
for times of strong background flaring that sometimes occur in \textit{XMM-Newton}
data, and two bursts were removed from the 2009 data. Details of each
of the four observations, including the total pn exposure time after
removing the bad time intervals, are listed in Table~\ref{tab:Obs}.

\section{Imaging Analysis and Results}

In order to search for and characterize any extended emission around
1E 1547.0$-$5408, we began by removing all point sources detected in
the field of each observation other than the magnetar, as well as
the out-of-time pn events present in the 2009 observation. To construct
a radial profile, we extracted events from concentric annuli having
width $2\arcsec$ centered on the position of the star, as determined
by a standard centroid search algorithm. The number of counts in each
bin was divided by both the geometric area of each extraction region
and the mean exposure time therein as determined from the unvignetted
exposure map. This procedure corrects for the chip gaps, dead pixels,
and removed regions on the detector.

The radial profile of a point source in \textit{XMM-Newton} is given
by the energy-dependent, radially averaged point-spread function (PSF).
Using the SAS task \texttt{eradial} we extracted from instrument calibration
files the theoretical PSF at 1\,keV energy intervals from 1 to 12
keV for each observation. These component PSFs were weighted such
that the PSF at energy $E_{i}$ was given the weight $W(E_{i})=\frac{N_{10}\left(\left|E-E_{i}\right|<0.5\,\mathrm{keV}\right)}{N_{10}\left(E\right)}$,
where $N_{10}(E)$ is the total number of counts within $10\arcsec$
of the source position. They were then summed to produce a weighted
PSF, $S(r)$. Finally, by scaling $S(r)$ using the formula \[
P(r)=a\cdot S(r)+b,\]
we created an expected point-source radial profile $P(r)$ for each
observation. Here, $b$ is a spatially uniform background count rate
found by averaging the count rate in all the bins of the observed
profile a sufficient distance away from the source. The normalization
factor $a$ was derived via a least-squares fit of $P(r)$ to the
first five bins $\left(10\arcsec\right)$ of the observed radial profile,
under the assumption that the contribution of any extended emission
to the profile in that region would be minimal. We found, however,
that the least-squares fit tended to be poor, especially for the observations
with higher count rate and thus better statistics. This suggested
that the uncertainty on $a$ based on the $\chi^{2}$ was not reliable
because of additional systematic errors affecting the fit. For example,
in addition to possible contamination by extended emission, the $2\arcsec$
size chosen for the radial bins oversamples the $4\farcs1$ square
pixels of the pn detector. Therefore, in order to better estimate
the uncertainty, we obtained a range of possible values for $a$ by
making a least-squares fit of $P(r)$ to any four of the first five
and any five of the first six bins of the observed profile. Our uncertainty
estimate was then given by $\delta a=(a_{\mathrm{max}}-a_{\mathrm{min}})/2$.

\begin{figure}
\hfill{}\includegraphics[width=0.4\textwidth]{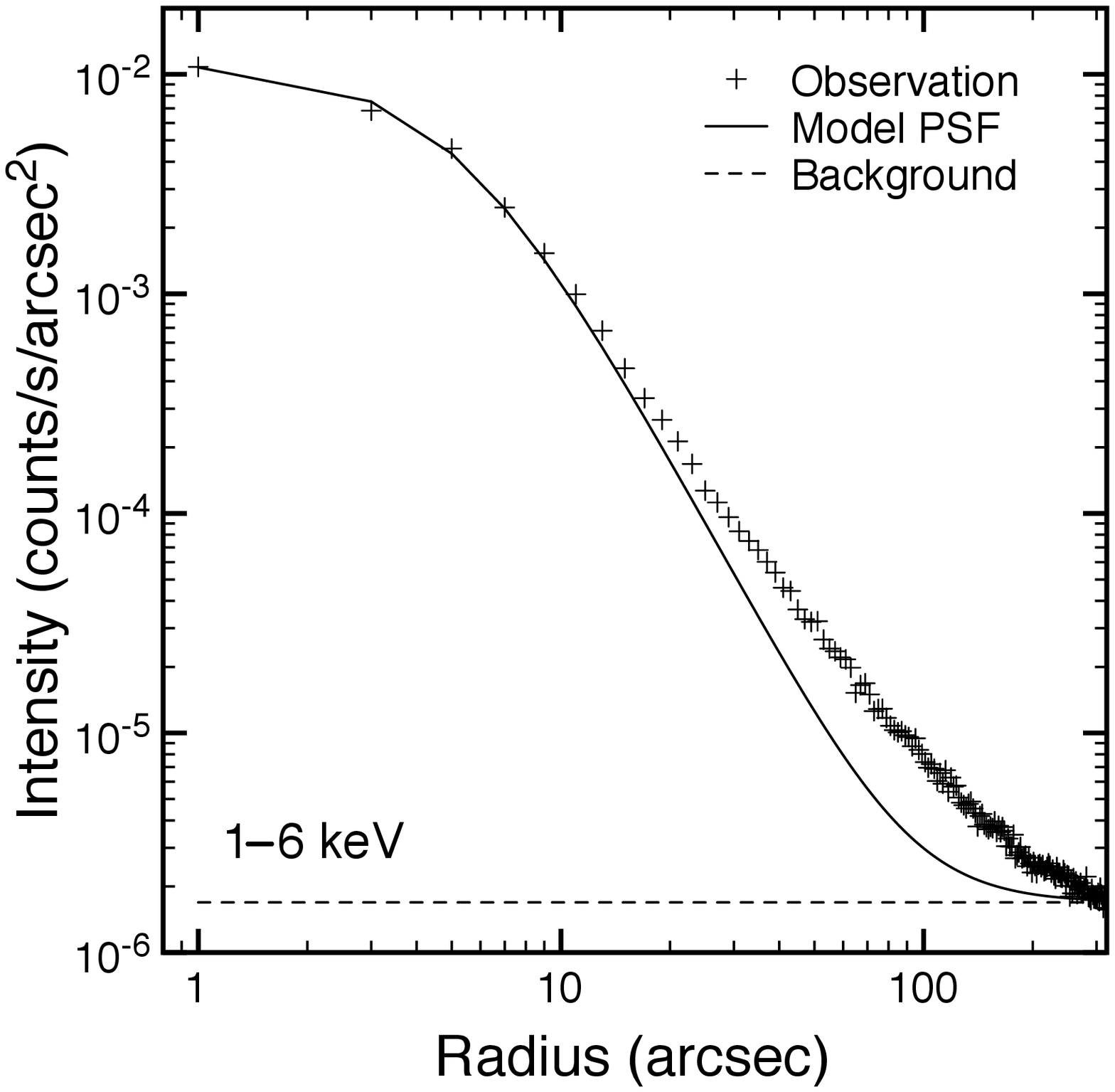}\qquad{}\includegraphics[width=0.4\textwidth]{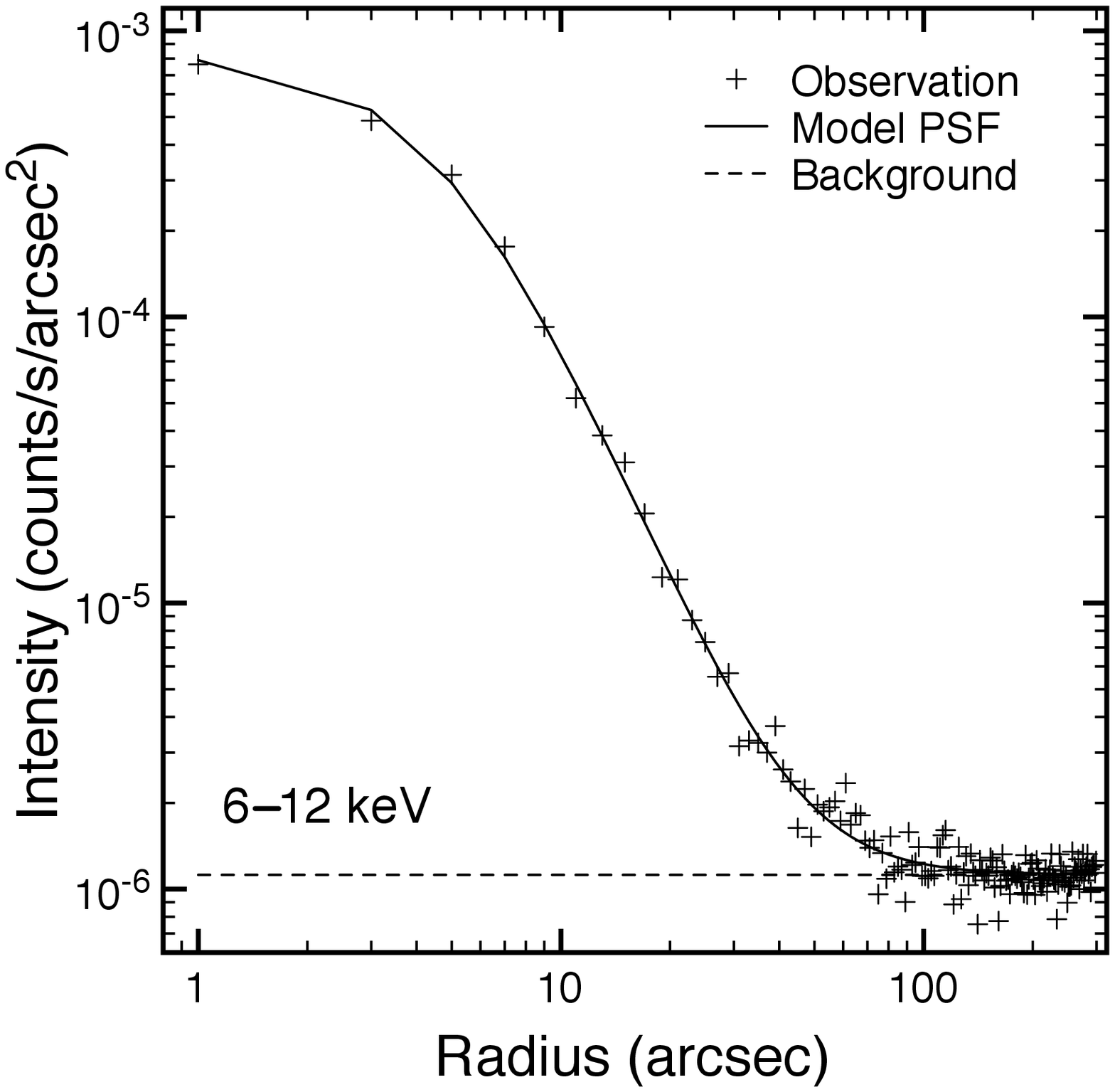}\hfill{}

\caption{\label{fig:RProf10}Left: 2010 February radial profile of 1E~1547.0$-$5408
in the 1--6\,keV energy band. Right: same but for 6--12\,keV.}

\end{figure}

The 2010 radial profiles of 1E 1547.0$-$5408 in the 1--6\,keV and
6--12\,keV energy bands are shown in Figure~\ref{fig:RProf10}.
Below 6\,keV the observed profile has a significant excess of counts
over the expected point-source profile, extending out to $r\approx4\farcm5$;
conversely, above 6\,keV no excess is detected and the observed profile
is consistent with a point source. Radial profiles constructed from
the three archival data sets display similar results, although the
shape and extent of the excess vary among the observations, and the
three dust-scattering rings reported in \citet{tve+10} are visible
in the 2009 data. We therefore confirm the presence of extended emission
around 1E 1547.0$-$5408 below 6\,keV as previously reported by VB09
and \citet{tve+10}. We also find that the extended emission is brighter
at low energies ($<$3\,keV), in line with the soft spectrum reported by
VB09.

In Figure~\ref{fig:Ext}, we plot the 1--6\,keV count rate of the
extended emission, $I_{\mathrm{ext}}$, as a function of the total
background-subtracted point-source count rate, $I_{\mathrm{ps}}$,
for two regions: region A, an annulus centered on the source with
radius $20\arcsec<r<40\arcsec$; and region B, a similar annulus
but with radius $40\arcsec<r<150\arcsec$. In both regions we find
a tight correlation between the two quantities; in fact, the first
three points (2006, 2007, and 2010) fit well to a straight line, although
the fourth point (2009) lies above the extrapolated linear fit in
both regions A and B (but see Section~\ref{sub:Dust}). The extended
emission flux varies wildly between observations, increasing and decreasing
along with the flux of the pulsar. For example, in region A the extended
emission brightened by a factor of nearly 50 between the 2006 and
2009 observations, and by the following year it had faded by almost
a factor of three.

\begin{figure}
\hfill{}\includegraphics[height=0.4\textheight]{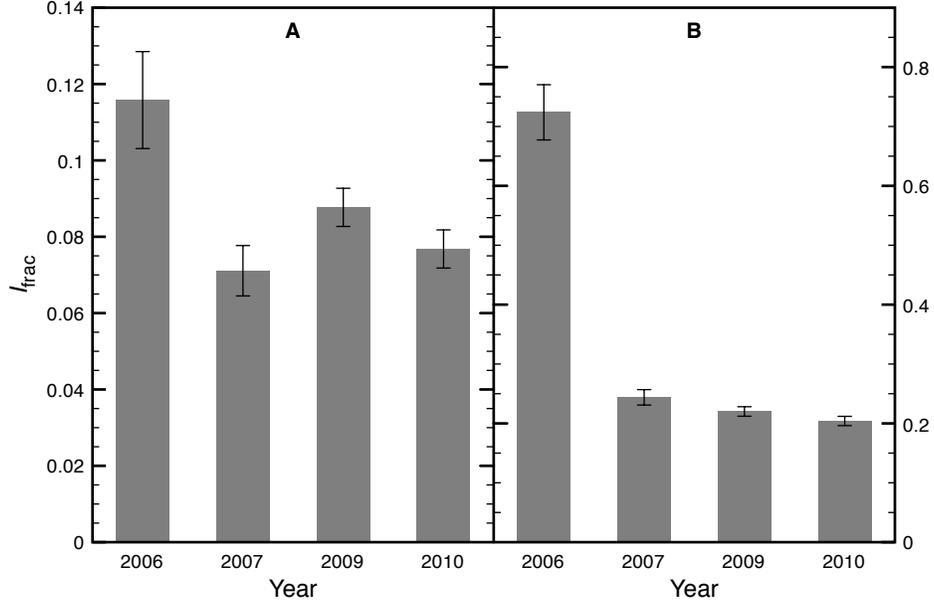}\hfill{}

\caption{\label{fig:IFrac}Fractional intensity of the extended emission, $I_{\mathrm{frac}}=I_{\mathrm{ext}}/I_{\mathrm{ps}}$,
in the 1--6\,keV energy band for regions A and B of all four
\textit{XMM-Newton} observations of 1E~1547.0$-$5408.}

\end{figure}

Figure~\ref{fig:IFrac} shows the fractional intensity of the extended
emission, $I_{\mathrm{frac}}=I_{\mathrm{ext}}/I_{\mathrm{ps}}$ in
regions A and B for all four observations of 1E 1547.0$-$5408. The
most prominent feature in both regions is that $I_{\mathrm{frac}}$
in 2006 is notably higher than in the other three observations, particularly
in region B.

\subsection{Spectral Analysis\label{sub:Spect}}

Because of contamination from the broad wings of the \textit{XMM-Newton}
PSF, we could not simply extract spectra of the extended emission
from regions A and B. For example, in each observation, less than
half of the total background-subtracted counts found in region A were
contributed by the extended source. Therefore, we instead computed
a simple hardness ratio for the extended emission, $\mathrm{HR_{ext}}\equiv I_{\mathrm{ext}}\left(\textrm{3--6\,\ keV}\right)/I_{\mathrm{ext}}\left(\textrm{1--3\,\ keV}\right)$,
and we list in Table~\ref{tab:HR}, for all four observations, these
hardness ratios in regions A and B. For comparison, the table also
lists the hardness ratio of the point source, $\mathrm{HR_{ps}}\equiv I_{\mathrm{tot}}\left(\textrm{3--6\,\ keV}\right)/I_{\mathrm{tot}}\left(\textrm{1--3\,\ keV}\right)$,
where $I_{\mathrm{tot}}$ is the background-subtracted count rate
within $10\arcsec$ of the source position.

\begin{figure}
\hfill{}\includegraphics[height=0.4\textheight]{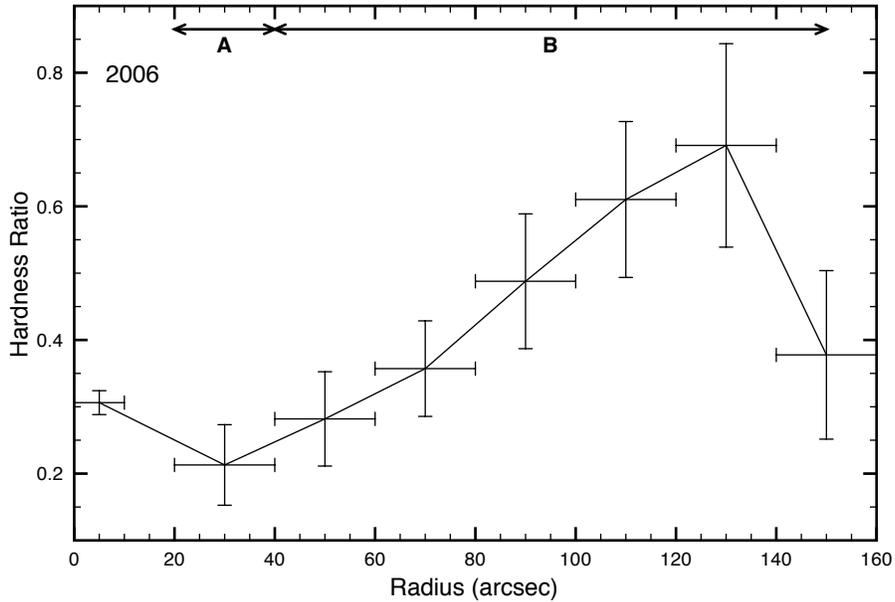}\hfill{}

\caption{\label{fig:Hard}Hardness ratio, $\mathrm{HR}\equiv I_{\mathrm{ext}}\left(\textrm{3--6\,\ keV}\right)/I_{\mathrm{ext}}\left(\textrm{1--3\,\ keV}\right)$,
for the extended emission of 1E~1547.0$-$5408 in 2006. The first bin
gives the hardness ratio of the point source, whereas all subsequent
bins are for the extended emission only. Regions A and B are labeled.}

\end{figure}

We find that in region A, $\mathrm{HR_{ext}<HR_{ps}}$ for all four
observations, meaning that the extended emission has a softer spectrum
than the point source, although in 2006 the hardness ratio is smaller
by only $1.5\sigma$. In region B, the results are the same as above
for all but the 2006 observation, where the extended emission spectrum
is instead harder than the source spectrum. This behavior is further
illustrated in Figure~\ref{fig:Hard}, which shows the 2006 hardness
ratio of the extended emission as a function of distance from the
source, beginning at $20\arcsec$ (the first bin provides the hardness
ratio of the point source itself).

Finally, assuming an absorbed power-law spectrum, the hardness ratios
in Table~\ref{tab:HR} can be used to constrain the photon index
$\Gamma$ and the flux of the extended emission. For example, taking
$N_{\mathrm{H}}=2.75\times10^{22}\,\mathrm{cm^{-2}}$ as in VB09,
we find that the 2--10\,keV unabsorbed flux of the extended emission
in region A varied from a minimum of $\sim$$4\times10^{-14}\,\mathrm{erg\, s^{-1}\, cm^{-2}}$
(for $\Gamma\approx4$) in 2006 to a maximum of $\sim$$3\times10^{-12}\,\mathrm{erg\, s^{-1}\, cm^{-2}}$
(for $\Gamma\approx3$) in 2009.

\section{Discussion}

Our analysis has confirmed the presence of extended emission around
1E 1547.0$-$5408, visible in four observations taken at very different
stages of its flux history. VB09 previously detected extended emission
around this source based primarily on a 2006 \textit{Chandra} observation
and interpreted it as a PWN and, farther out, the X-ray counterpart
of SNR G327.24--0.13. Here we reexamine this interpretation in light
of our new data.

\subsection{A Pulsar Wind Nebula?}

In the context of conventional rotation-powered pulsars, given the
low spin-down power of 1E 1547.0$-$5408, we generally would not expect
it to harbor a bright PWN in X-rays. For $\dot{E}=10^{35}$\,erg\,s$^{-1}$,
the typical X-ray efficiency of a PWN is about $10^{-4}$ \citep{kp08}.
Even allowing for the X-ray efficiency to be up to an order of magnitude
greater, this predicts an unabsorbed PWN flux of $\lesssim$$5\times10^{-14}(d/4\,\mathrm{kpc})^{-2}$\,erg\,s$^{-1}$\,cm$^{-2}$
in the 0.5--8\,keV range for 1E 1547.0$-$5408. On the other hand, the
putative PWN as suggested by VB09 has an unabsorbed flux of $\sim$$1\times10^{-12}$\,erg\,s$^{-1}$\,cm$^{-2}$
in the same band. Thus, if the extended emission is entirely
rotation-powered, this would require an unusually high X-ray efficiency of
over 1\%. This problem could be alleviated, however, by hypothesizing that
magnetic power could be contributing to the nebula, as has been suggested by
\citet{rmg+09} for RRAT 1819$-$1458.

However, another issue is the soft spectrum of the putative PWN. The
hardness ratios in Section~\ref{sub:Spect} suggest a power-law spectrum
with photon index $\Gamma\approx3\textrm{--}4$, a range that is consistent
with the value reported in VB09 $\left(\Gamma=3.4\pm0.4\right)$.
This is much softer than previously reported PWNe, which typically
have $\Gamma\approx1.5\textrm{--}2$ \citep{kp08}, although VB09
proposed that the discrepancy could be explained as being somehow
a result of the magnetar nature of 1E 1547.0$-$5408.

\begin{figure}
\hfill{}\includegraphics[width=0.75\textwidth]{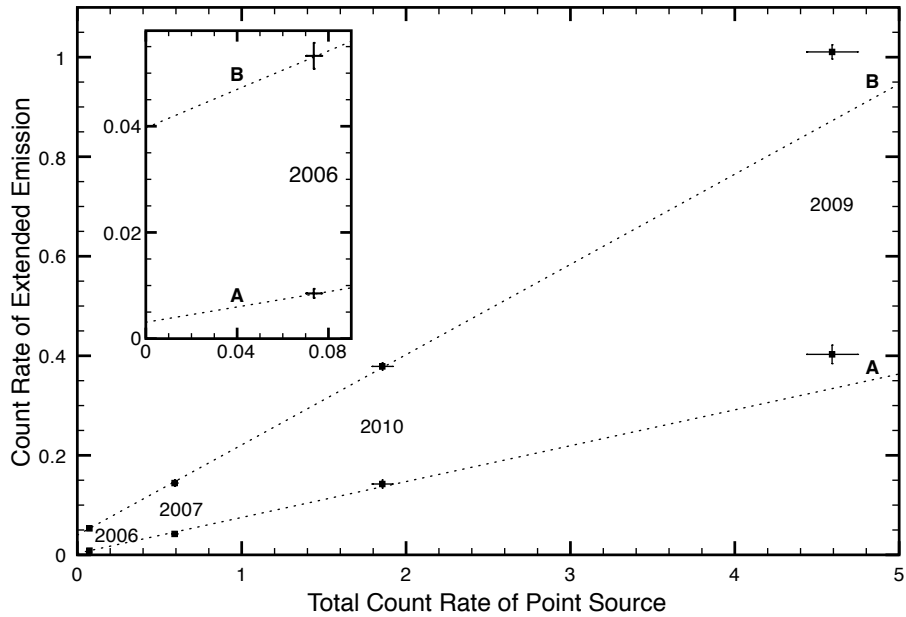}\hfill{}

\caption{\label{fig:Ext}1--6\,keV count rate of the extended emission integrated
over annular regions with $20\arcsec<r<40\arcsec$ (lower set of points,
labeled A) and $40\arcsec<r<150\arcsec$ (upper set of points, labeled
B) vs.\ the total background-subtracted 1--6\,keV count rate of
the point source for each of the four \textit{XMM-Newton} observations
of 1E~1547.0$-$5408. The dotted lines are fit to the left-most three
points. Inset: blow-up of the region near the origin covering the
2006 data points.}

\end{figure}

A more pressing problem with the PWN interpretation is its failure
to explain the strong flux correlation seen in Figure~\ref{fig:Ext}.
First, PWNe in rotation-powered pulsars have not been seen to have
large luminosity variations as are observed for 1E 1547.0$-$5408.
Even if energy injection due to outbursts played a role, we show here
that the observed fading time is incompatible with a synchrotron origin.
First, to estimate the magnetic field strength in the putative PWN,
we consider as an analogy the 2004 flare of SGR 1806$-$20, which
was accompanied by nebular radio emission. In that event, a total
energy of $2\times10^{46}$\,erg was released \citep{pbg+05}. \citet{l06}
proposed that the event could produce relativistic, weakly baryon-loaded
magnetic clouds analogous to a solar coronal mass ejection, and deduced
a total energy of $8\times10^{44}$\,erg for the relativistic electrons
plus magnetic field, with an average magnetic field $B$ of 0.1\,G within
a radius $\sim$$1.5\times10^{16}$\,cm. As the nebula expands, the
field strength decays with the volume $V$ as $V^{-1/2}$ or $V^{-2/3}$,
depending on whether the field is tangled \citep{gkg+05}. Assuming
the former case, we have scaled these values to those appropriate for the
properties of 1E 1547.0$-$5408 and find a conservative $B$-field
estimate of $<$$80\,\mu$G at $30\arcsec$, which corresponds to $1.8\times10^{18}$\,cm
from the source. We note that the $B$-field is much lower for the
latter field decay case. The synchrotron cooling timescale is then
$\tau_{{\rm syn}}=37(B/1\,\mu\mathrm{G})^{-3/2}(\varepsilon_{\gamma}/1\,\mathrm{keV})^{1/2}\,\mathrm{kyr\gtrsim120\, yr}$
for particles emitting at $\varepsilon_{\gamma}$=6\,keV. This is
incompatible with the flux decay timescale observed and shown in Figure~\ref{fig:Ext},
in particular between 2009 and 2010.

The above theory suggests that the particle energy was about 4\% of
the burst fluence for SGR 1806$-$20. The total burst fluence for the 2009
event of 1E 1547.0$-$5408 was measured to be $\gtrsim$$5\times10^{43}(d/10\,\mathrm{kpc})^{2}$\,erg
by \citet{mgw+09}, and estimated to be in the range $~10^{44}\textrm{--}10^{45}$\,erg
by \citet{tve+10}. Taking the uppermost value of $10^{45}$\,erg would give
an injected particle energy of $4\times10^{43}$\,erg. Assuming this results
in synchrotron emission from the radio regime up to 6\,keV, with a typical
photon index of 1.5, we find a power of $1.1\times10^{32}\,\mathrm{erg\, s^{-1}}$,
corresponding to a flux of $6\times10^{-14}(d/4\,\mathrm{kpc})^{-2}\,\mathrm{erg\, s^{-1}\, cm^{-2}}$
between 1 and 6\,keV. This is well below the observed extended emission flux
in 2009.

For completeness, we note that there are alternative models proposing
baryonic outflows for the magnetar outbursts \citep{gle+05,grt+06}.
If this is the case, then no detectable synchrotron X-rays are expected.

\subsection{Dust-scattering Halo\label{sub:Dust}}

Extended emission around an X-ray source can be produced by the scattering
of X-rays off dust particles between the source and observer. The
flux of such a dust-scattering halo is expected to be proportional
to the source flux \citep{ml91}. Returning to Figure~\ref{fig:Ext},
then, for a dust-scattering halo, all of the points should fit well
to a straight line (allowing for some scatter because the source spectrum
did not remain constant). This is indeed the case for the first three
points, although the 2009 point lies above the linear fit. In order
to fit with the linear trend, the 2009 source flux would have to be
15\%--20\% higher than what we observed. However, because the scattered
photons in a dust halo travel a longer path than photons observed
directly from the source, the halo flux depends in a complicated manner
on the recent history of the source flux over a period of hours or
days \citep{mg86}. The 2009 \textit{XMM} observation of 1E 1547.0$-$5408
was taken only 13 days after its January outburst, at which
time the magnetar's flux was decaying following a power law of index
$\alpha=-0.34$ to $-3.1$ \citep{bis+11,sk11}. As a result, the
source would have been bright enough to produce the observed halo
4.5--5.5 days prior to the observation, not an unreasonable timescale
for the evolution of a dust halo. We therefore conclude that the observed
variability in the extended emission flux is entirely consistent with
a dust-scattering halo.

VB09 rejected the dust-scattering halo interpretation of the extended
emission based on two arguments. The first one is that the extended
emission around 1E 1547.0$-$5408 had a harder spectrum than the source
itself. A dust halo, on the other hand, is expected to have a softer
spectrum than the source because the scattering cross section has
an inverse-square dependence on energy. From the hardness ratios in
Table~\ref{tab:HR} we find that, contrary to the claim by VB09,
the extended source in region A has an unambiguously softer spectrum
than the magnetar in 2007, 2009, and 2010, supporting the interpretation
of dust scattering. In 2006, the hardness ratios suggest a softer
spectrum too, although the difference is not statistically significant.
For comparison, though, VB09 reported that the photon index $\Gamma$
of the `PWN' differed from that of the point source by only $1\sigma$,
which is not statistically significant either. We cannot, therefore,
conclude that the spectrum of the extended emission in region A supports
either interpretation in 2006. It should be noted, however, that be
it harder or softer than the point source, the extended emission in
2006 still has a much softer spectrum than any previously reported
PWN, as discussed above.

Table~\ref{tab:HR} also indicates similar results for region B as
in region A. The extended emission shows a softer spectrum than the
magnetar in all observations except 2006. Again, we note that although
the extended emission spectrum is harder than the point source in
2006, it is still very soft overall.

Our investigations so far strongly support that the extended emission
observed around 1E 1547.0$-$5408 in regions A and B is dominated by
a dust-scattering halo, at least in 2007, 2009, and 2010, although
this interpretation is less clear in 2006. We now examine the other
argument given in VB09 against dust scattering: that the extended
emission was too bright, especially above 3\,keV, to be a dust halo.
They estimated the expected fractional halo intensity and the dust-scattering
optical depth $\tau_{\mathrm{sca}}$ based on models by \citet{ps95}
and \citet{d03}, which depend on the absorption column $N_{\mathrm{H}}$,
the X-ray energy $E$, and a parameter $\beta$ describing the distribution
of the dust between source and observer. Following a similar procedure,
we assumed an effective energy $E=2\,\mathrm{keV}$ for the 1--6\,keV
photons based on the spectrum of the magnetar, took $\beta=1$ (meaning
most of the dust is close to the source) and $\tau_{\mathrm{sca}}=1.5$,
as in VB09, and calculated $I_{\mathrm{frac}}$ for regions A and
B. This gives $I_{\mathrm{frac}}=0.08$ for region A and $I_{\mathrm{frac}}=0.18$
for region B, which are in a good agreement%
\footnote{In region B, the calculated $I_{\mathrm{frac}}$ is actually below
the observed value by $3\sigma$. However, if we subtract the non-dust
contribution to the extended emission from the SNR (see below and
Table~\ref{tab:NonDust}) then the two values are in full agreement.%
} with the 2010 values in Figure~\ref{fig:IFrac}, suggesting that
dust scattering is adequate to explain the brightness of the extended
emission, at least in 2010. For pure dust scattering we expect $I_{\mathrm{frac}}$
in each region to be the same for each observation, except in 2009
where it should be higher due to source variability as discussed above.
Indeed, there is a good agreement between the 2007, 2009, and 2010
observations, but the 2006 value of $I_{\mathrm{frac}}$ stands out.
In region A, it is $3\sigma$ higher than what is expected from the
2007 and 2010 data, and in region B an even larger increase is evident,
with $I_{\mathrm{frac}}$ being $>$$10\sigma$ higher in 2006 than in
any subsequent observation.

The best explanation for all of our results is that the extended emission
around 1E 1547.0$-$5408 consists of a dust-scattering halo plus an
additional component independent of the source flux. This secondary
component is significant mainly in region B and becomes noticeable
only when the halo is faint, as is the case for the 2006 observation.
In order to better quantify it we return to Figure~\ref{fig:Ext}.
For pure dust scattering, the linear fits in the diagram should pass
through the origin. As seen from the inset, however, both fits have
a positive $y$-intercept, suggesting that some of the extended emission
does not come from the dust halo. In Table~\ref{tab:NonDust}, we
list the values of the $y$-intercepts for regions A and B and the extended
emission count rate in 2006, and we calculate the fraction of the
latter that was not contributed by dust scattering.

We find that in region B, $75\%\pm6\%$ of the extended emission in
2006 is not from the dust halo. Since our region B mostly corresponds
to the SNR region $\left(45\arcsec<r<174\arcsec\right)$ from VB09
and therefore to the location of the radio SNR shell, we conclude
that the X-ray counterpart of this shell is the source of the non-dust
extended emission here. 

Region A corresponds roughly to the `PWN' region $\left(4\arcsec<r<45\arcsec\right)$
from VB09, noting that the broad PSF of \textit{XMM-Newton} restricts
us to $r>20\arcsec$. Unlike farther out, the 2006 extended emission
in region A is still dominated by the dust halo; only $36\%\pm12\%$
of it comes from another source. In fact, since the significance is
only $3\sigma$ above zero, it is possible that dust scattering alone
is sufficient to explain all of the extended emission in region A.
Nevertheless, we can use the parameters in Table~\ref{tab:NonDust}
to estimate an upper limit on the flux of a possible PWN, assuming
an absorbed power-law spectrum with $N_{\mathrm{H}}=2.75\times10^{22}\,\mathrm{cm^{-2}}$,
as in VB09, and a photon index of $\Gamma=2$, as is typical of PWNe.
We find a $3\sigma$ upper limit%
\footnote{In this case, the most stringent upper limit found was based on the
non-detection of extended emission above 6\,keV, not the parameters
from Table~\ref{tab:NonDust}.%
} on the 2--10\,keV unabsorbed flux of $\lesssim$$4.7\times10^{-14}\,\mathrm{erg\, s^{-1}\, cm^{-2}}$,
corresponding to a luminosity of $9\times10^{31}(d/4\,\mathrm{kpc})^{2}\,\mathrm{erg\, s^{-1}}$.
This implies an X-ray efficiency of $L_{\mathrm{X}}/\dot{E}\lesssim9\times10^{-4}(d/4\,\mathrm{kpc})^{2}$.
We also repeated our estimation for a softer PWN spectrum of $\Gamma=3$,
but found that the 2--10\,keV upper limit was largely insensitive to
changes in $\Gamma$.

\section{Conclusions}

In this paper, we have examined multi-epoch \textit{XMM-Newton} data
for the magnetar candidate 1E 1547.0$-$5408 and we show that the
observed extended emission surrounding the source is dominated by
dust-scattered magnetar emission. Specifically we find that the luminosity
of the nebular emission is proportional to the source flux, as expected
for dust scattering, but not seen in any known PWN or other magnetar
candidate. Additional strong evidence for dust-scattering comes from
spectral and energetics arguments, as well as from the disagreement
between the observed nebular variability time scale and the expected
synchrotron loss time in the PWN interpretation. We note that contrary to
a previous claim (VB09), even in 2006 when the source was relatively faint,
$64\%\pm12\%$ of the nebular emission is from dust scattering. We cannot,
however, rule out the presence of a faint PWN with luminosity $\lesssim$$9\times10^{31}(d/4\,\mathrm{kpc})^{2}\,\mathrm{erg\, s^{-1}}$
in the 2--10 keV band, a limit three times lower than the previously claimed
detection ($\sim$$2.9\times10^{32}\,\mathrm{erg\, s^{-1}}$ from VB09).
Deep observations of this source when the magnetar is in quiescence are
necessary to test this hypothesis. We do, on the other hand, find strong
evidence for non-dust-scattered extended X-ray emission at angular distance
$\sim$$40\arcsec\textrm{--}150\arcsec$, which we argue is from the SNR
shell surrounding the pulsar, as previously reported by VB09.

With the absence of evidence for a PWN surrounding any AXP or SGR,
now including 1E 1547.0$-$5408, previous models for the production
of PWNe by magnetars \citep{tb98,hck99} remain unsupported. On the
other hand with clear evidence for the existence of PWNe surrounding
many high-magnetic-field radio pulsars (e.g., PSR J1846--0258: \citealp{hcg03};
PSR J1119--6127: \citealp{gs03}) which generally have substantially
higher spin-down luminosities than any known AXP or SGR, the production
of the relativistic particle wind necessary to generate an observable
PWN seems intimately tied to the rotation-derived power, rather than
that from magnetic-field decay. This then makes the detection of the
surprisingly bright X-ray PWN surrounding the presumably rotation-powered
but relatively high-$B$ RRAT PSR J1819$-$1458 \citep{rmg+09} particularly
interesting and worthy of follow-up.

\acknowledgements{This research is based on observations obtained with \textit{XMM-Newton},
an ESA science mission with instruments and contributions directly
funded by ESA Member States and NASA. V.M.K. receives support from
NSERC via a Discovery Grant, FQRNT via the Centre de Recherche Astrophysique
du Qu\'ebec, CIFAR, a Killam Research Fellowship, and holds a Canada
Research Chair and the Lorne Trottier Chair in Astrophysics and Cosmology.
C.-Y.N. is a CRAQ postdoctoral fellow and a Tomlinson postdoctoral
fellow.}

\begin{deluxetable}{ccccc}
\tablewidth{0pt}
\tablecolumns{5}
\tablecaption{Summary of \textit{XMM-Newton} Observations of 1E~1547.0$-$5408\label{tab:Obs}}
\tablehead{\colhead{Date} & \colhead{Obs ID} & \colhead{Exposure (ks)} & \colhead{Count Rate (cnt\,s$^{-1}$)\tablenotemark{a}} & \colhead{Mode/Filter\tablenotemark{b}}}

\startdata
2006 Aug 21 & 0402910101 & 38.7 & 0.074 & FF/Medium \\
2007 Aug 9 & 0410581901 & 11.6 & 0.59 & LW/Medium \\
2009 Feb 3 & 0560181101 & 48.9 & 4.6 & FF/Thick \\
2010 Feb 10 & 0604880101 & 35.7 & 1.9 & LW/Medium \\
\enddata

\tablenotetext{a}{Background-subtracted count rate of the point source in the 1--6 keV energy range.}

\tablenotetext{b}{The time resolutions of the operating modes of the EPIC pn camera are: Full Frame (FF): 73.4 ms; Large Window (LW): 47.7 ms.}
\end{deluxetable}

\begin{deluxetable}{cccc}
\tablewidth{0pt}
\tablecolumns{4}
\tablecaption{Hardness Ratios for 1E~1547.0$-$5408 and the Surrounding Extended Emission\label{tab:HR}}
\tablehead{\colhead{Observation} & \multicolumn{3}{c}{Hardness Ratio\tablenotemark{a}} \\
\colhead{} & \colhead{Point Source} & \colhead{Region A} & \colhead{Region B}}

\startdata
2006 & 0.306(18) & 0.21(6) & 0.46(5) \\
2007 & 0.482(16) & 0.18(5) & 0.32(4) \\
2009 & 1.014(6)\phn & 0.55(4) & 0.45(1) \\
2010 & 0.814(8)\phn & 0.34(4) & 0.37(2) \\
\enddata

\tablenotetext{a}{The hardness ratio is defined as $I\left(\textrm{3--6\,keV}\right)/I\left(\textrm{1--3\,keV}\right)$, where $I$ is the background-subtracted count rate of the point source or extended emission. Numbers in parentheses are $1\sigma$ uncertainties.}
\end{deluxetable}

\begin{deluxetable}{cccc}
\tablewidth{0pt}
\tablecolumns{4}
\tablecaption{Contribution to Extended Emission Not From Dust-scattering\label{tab:NonDust}}
\tablehead{\colhead{Region} & \multicolumn{2}{c}{Count Rate of Extended Emission (cnt\,s$^{-1}$)\tablenotemark{a}} & \colhead{Fraction Not Attributable} \\
\colhead{} & \colhead{Extrapolated $y$-intercept\tablenotemark{b}} & \colhead{Observed Value in 2006} & \colhead{to Dust Scattering\tablenotemark{a}}}

\startdata
A & 0.003(1) & 0.0085(4) & 0.36(12) \\
B & 0.040(3) & 0.053(4) & 0.75(6) \\
\enddata

\tablenotetext{a}{Numbers in parentheses are $1\sigma$ uncertainties.}

\tablenotetext{b}{Extrapolated $y$-intercept for the linear fits in Figure~\ref{fig:Ext}.}

\end{deluxetable}

\end{document}